\documentclass{article}

\usepackage{multirow}
\usepackage{float}

\usepackage[utf8]{inputenc} 
\usepackage[T1]{fontenc}    
\usepackage{url}            
\usepackage{booktabs}       
\usepackage{amsfonts}       
\usepackage{nicefrac}       
\usepackage{microtype}      

\usepackage{mathtools}
\usepackage{natbib}
\usepackage{xcolor}
\usepackage{url}
\usepackage{amsmath}
\usepackage{amssymb}
\usepackage{graphicx}
\usepackage{subcaption}
\usepackage{nicefrac}
\usepackage{appendix}
\usepackage{enumitem}
\definecolor{darker}{rgb}{0,0.15,0.7}
\usepackage[colorlinks, urlcolor=darker, citecolor=darker, linkcolor=darker]{hyperref} 
\setcitestyle{round}


\usepackage{algorithm}
\usepackage{algorithmic}

\usepackage{amsthm}

\begingroup
    \makeatletter
    \@for\theoremstyle:=definition,remark,plain\do{%
        \expandafter\g@addto@macro\csname th@\theoremstyle\endcsname{%
            \addtolength\thm@preskip\parskip
            }%
        }
\endgroup

\newcommand{\vv}[1]{\boldsymbol{#1}}

\newcommand{\NA}{---}

\usepackage[accepted]{icml2020}

\icmltitlerunning{WaveFlow: A Compact Flow-based Model  for Raw Audio}

\begin{document}

\twocolumn[
\icmltitle{WaveFlow: A Compact Flow-based Model  for Raw Audio}




\begin{icmlauthorlist}
\icmlauthor{Wei Ping}{to}~\quad
\icmlauthor{Kainan Peng }{to}\ \
\icmlauthor{Kexin Zhao }{to}\ \
\icmlauthor{Zhao Song}{to}
\end{icmlauthorlist}
\icmlaffiliation{to}{Baidu Research, 1195 Bordeaux Dr, Sunnyvale, CA. \\
Project page:~\url{https://waveflow-demo.github.io/}}
\icmlcorrespondingauthor{Wei Ping}{weiping.thu@gmail.com}
\icmlkeywords{Flow-based model, generative models, raw waveform synthesis, neural text-to-speech}

\vskip 0.3in
]



\printAffiliationsAndNotice{}  

\begin{abstract}
In this work, we propose WaveFlow, a small-footprint generative flow for raw audio, which is directly trained with maximum likelihood.
It handles the long-range structure of 1-D waveform with a dilated 2-D convolutional architecture, while modeling the local variations using expressive autoregressive functions.
WaveFlow provides a unified view of likelihood-based models for 1-D data, including WaveNet and WaveGlow as special cases.
It generates high-fidelity speech as WaveNet, while synthesizing several orders of magnitude faster as it only requires a few sequential steps to generate very long waveforms with hundreds of thousands of time-steps. Furthermore, it can significantly reduce the likelihood gap that has existed between autoregressive models and flow-based models for efficient synthesis.
Finally, our small-footprint WaveFlow has only 5.91M parameters, which is 15$\times$ smaller than WaveGlow. It can generate 22.05~kHz high-fidelity audio 42.6$\times$ faster than real-time~(at a rate of 939.3 kHz) on a V100 GPU without engineered inference kernels.
\end{abstract}

\section{Introduction}
Deep generative models have obtained noticeable successes for modeling raw audio in high-fidelity speech synthesis and music generation~\citep[e.g.,][]{oord2016wavenet, dieleman2018challenge}. 
Autoregressive models are among the best performing generative models for raw waveforms, providing the highest likelihood scores and generating high-fidelity audios~\citep{oord2016wavenet,mehri2016samplernn, kalchbrenner2018efficient}. 
One of the most successful examples is WaveNet~\citep{oord2016wavenet}, an autoregressive model for waveform synthesis.
It operates at the high temporal resolution~(e.g., 24 kHz) of raw audio and sequentially generates 1-D waveform samples at inference. 
As a result, WaveNet is prohibitively slow for speech synthesis and one has to develop highly engineered kernels~\citep{arik2017DV1, nv-wavenet18} for real-time inference, which is a requirement for most production text-to-speech systems.

Flow-based models~\citep{dinh2014nice, rezende2015variational} are a family of generative models, in which a simple initial density is transformed into a complex one by applying a series of invertible transformations.
One group of models are based on \emph{autoregressive transformation}, including {autoregressive flow}~(AF) and {inverse autoregressive flow}~(IAF) as the ``dual'' of each other~\citep{kingma2016improved, papamakarios2017masked, huang2018neural}.
AF is analogous to autoregressive models, which performs parallel density evaluation and sequential synthesis. 
In contrast, IAF performs parallel synthesis but sequential density evaluation, making likelihood-based training very slow.
Parallel WaveNet~\citep{oord2017parallel} distills an IAF from a pretrained autoregressive WaveNet, which gets the best of both worlds. 
However, one has to apply the Monte Carlo method to approximate the intractable KL divergence in distillation.
Instead, ClariNet~\citep{ping2018clarinet} simplifies the probability density distillation by computing a regularized KL divergence in closed-form.
Both of them require a pretrained WaveNet teacher and a set of auxiliary losses for high-fidelity synthesis, which complicates the training pipeline and increases the cost of development.

Another group of flow-based models are based on \emph{bipartite transformation}~\citep{dinh2016density, kingma2018glow}, which provide likelihood-based training and parallel synthesis. 
Most recently, WaveGlow~\citep{prenger2019waveglow} and FloWaveNet~\citep{kim2018flowavenet} apply {Glow}~\citep{kingma2018glow} and {RealNVP}~\citep{dinh2016density} for waveform synthesis, respectively.
However, the bipartite flows require more layers, larger hidden size, and huge number of parameters to reach comparable capacities as autoregressive models. 
In particular, WaveGlow and FloWaveNet have 87.88M and 182.64M parameters with 96 layers and 256 residual channels, respectively, whereas a typical 30-layer WaveNet has 4.57M parameters with 128 residual channels.
Moreover, both of them \emph{squeeze} the time-domain samples on the \emph{channel} dimension before applying the bipartite transformation, which may lose the temporal order information and make them less efficient at modeling waveform sequence. 

In this work, we present WaveFlow, a compact flow-based model for raw audio, which features \emph{i}) simple training, \emph{ii}) high-fidelity \& ultra-fast synthesis, and \emph{iii}) small footprint. 
Specifically, we make the following contributions:
\vspace{-0.3em}
\begin{enumerate}[itemsep=-0.00pt, topsep=0pt, leftmargin=1.5em]
    \item WaveFlow is directly trained with maximum likelihood without probability density distillation and auxiliary losses as used in Parallel WaveNet~\citep{oord2017parallel} and ClariNet~\citep{ping2018clarinet}, which simplifies the training pipeline and reduces the cost of development.
     WaveFlow \emph{squeezes} the 1-D waveform samples into a 2-D matrix and processes the local adjacent samples with autoregressive functions without losing temporal order information.
    We implement WaveFlow with a dilated 2-D convolutional architecture~\citep{yu2015multi}, which leads to 15$\times$ fewer parameters, and faster synthesis speed than WaveGlow~\citep{prenger2019waveglow}. 
    \vspace{-.04cm}
    \item  WaveFlow provides a unified view of likelihood-based models for raw audio, which includes both WaveNet and WaveGlow as special cases and allows us to explicitly trade inference parallelism for model capacity.
    We systematically study these models in terms of test likelihood and audio fidelity.
    We demonstrate that a moderate-sized WaveFlow can obtain comparable likelihood and synthesize high-fidelity speech as  WaveNet, while synthesizing thousands of times faster. 
    In previous work, there is a large likelihood gap between autoregressive models and flow-based models which provide efficient sampling~\citep{ho2019flow++, tran2019discrete}.
    \vspace{-.04cm}
    \item  For practitioners, our small WaveFlow has only 5.91M parameters by utilizing the compact autoregressive functions for modeling local signal variations. It synthesizes 22.05 kHz high-fidelity speech (MOS: 4.32) more than 40$\times$ faster than real-time on a Nvidia V100 GPU. 
    In contrast, WaveGlow~\citep{prenger2019waveglow} requires  87.88M parameters for generating high-fidelity speech. 
    The small memory footprint is preferred in production TTS systems, especially for on-device deployment.
\vspace{-.2em}
\end{enumerate}
We organize the rest of the paper as follows. 
Section~\ref{sec:flow_based_models} reviews the flow-based models.
We present WaveFlow in Section~\ref{sec:waveflow} and discuss related work in Section~\ref{sec:related_work}.
We report experimental results in Section~\ref{sec:experiment} and discuss the pros and cons of different methods in Section~\ref{sec:conclusion}.

\vspace{-.2em}
\section{Flow-based generative models}
\label{sec:flow_based_models}
\vspace{-.1em}
Flow-based models~\citep{dinh2014nice,dinh2016density,rezende2015variational} transform a simple density $p(\vv z)$~(e.g., isotropic Gaussian) into a complex data distribution $p(\vv x)$ by applying a bijection $\vv x = f(\vv z)$, where $\vv x$ and $\vv z$ are both $n$-dimensional. 
The probability density of $\vv x$ can be obtained through the change of variables formula:
\begin{align}
\label{eq:change_of_variable}
    p(\vv x) =  p(\vv z)  
    \left| \det\left( \frac{\partial f^{-1}(\vv x) }{\partial \vv x} \right) \right| , 
\end{align}
where $\vv z = f^{-1}(\vv x)$ is the inverse of the bijection, 
and $\det \big(\frac{\partial f^{-1}(\vv x) }{\partial \vv x} \big)$ is the determinant of its Jacobian. 
In general, it takes $O(n^3)$ to compute the determinant, which is not scalable in high-dimension.
There are two notable groups of flow-based models with triangular Jacobians and tractable determinants, which are based on autoregressive and bipartite transformations, respectively.
Before delving into details, we summarize the model capacities and parallelisms of after-mentioned flow-based models in Table~\ref{tab:summary_flow}.

\vspace{-.2em}
\subsection{Autoregressive transformation}
\begin{figure*}[t] \centering
\begin{tabular}{cc}
\hspace{-.4cm}
\includegraphics[height=4.1cm, clip]{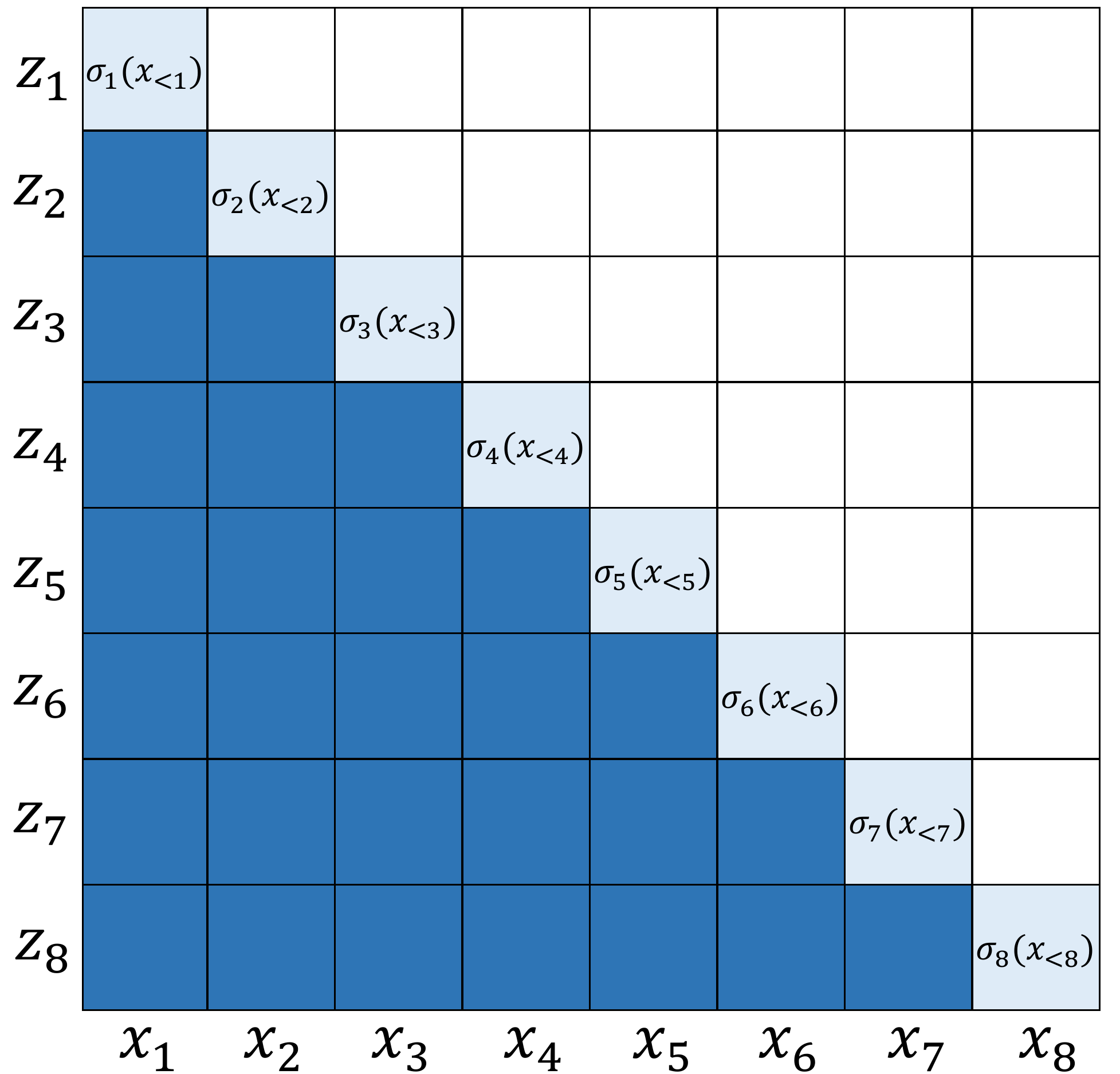} 
&
\hspace{1.3cm}
\includegraphics[height=4.1cm, clip]{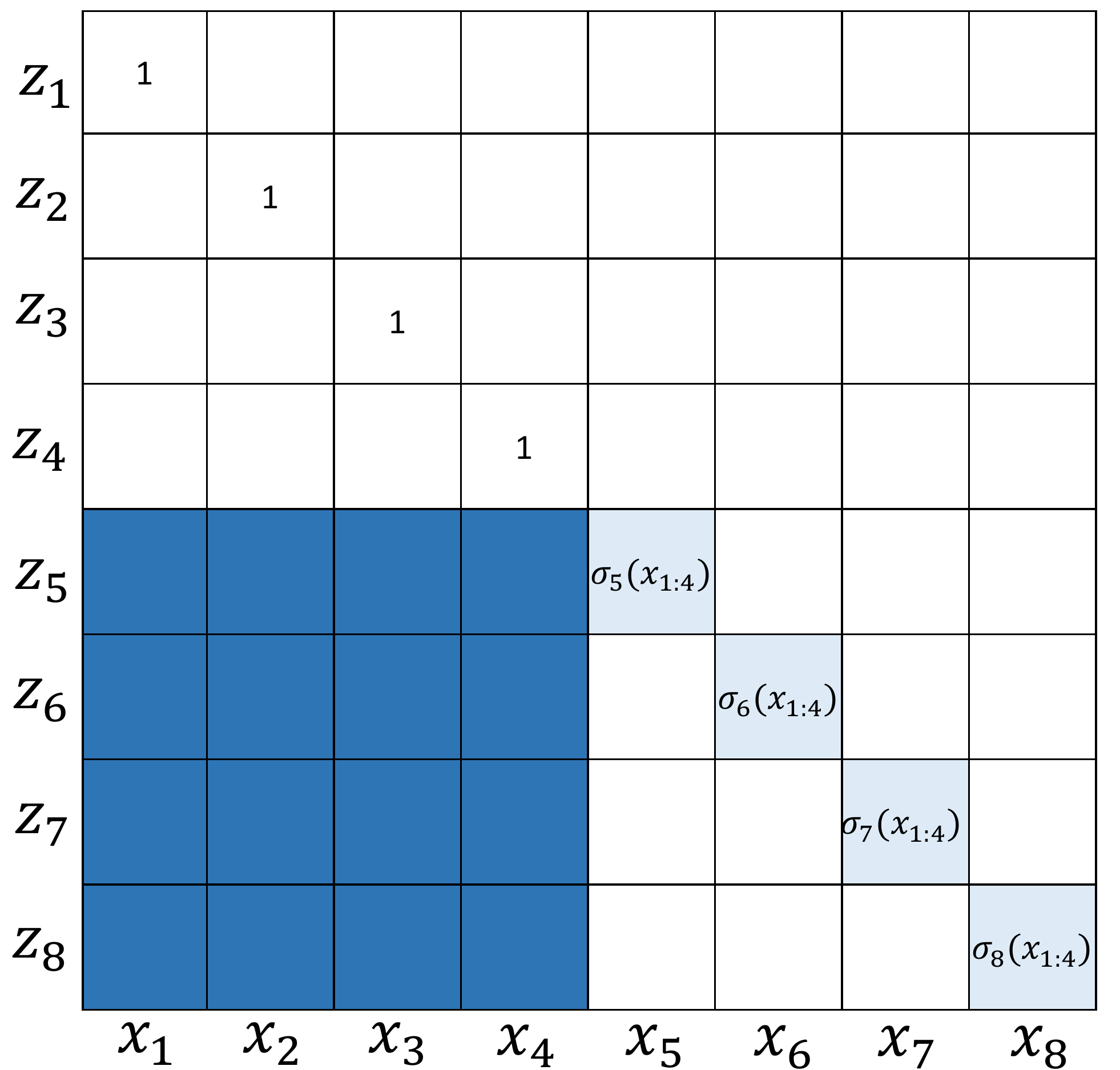} 
\\
\vspace{-.7em}
{\hspace{-.5em} \small (a)}  & \hspace{1.2cm} {\small(b)}
\\
\end{tabular}
\vspace{-.8em}
\caption{The Jacobian $\frac{\partial f^{-1}(\vv x) }{\partial \vv x}$ of (a) an autoregressive transformation, and (b) a bipartite transformation.
The blank cells are zeros and represent the independent relations between $z_i$ and $x_{j}$.
The light-blue cells with scaling variables $\sigma$ represent the linear dependencies.
The dark-blue cells represent complex non-linear dependencies.}
\vspace{-.7em}
\label{fig:Jacobian} %
\end{figure*}

\vspace{-.1em}
The autoregressive flow~(AF) and inverse autoregressive flow~(IAF)~\citep{kingma2016improved, papamakarios2017masked} use autoregressive transformations.
Specifically, AF defines $\vv z = f^{-1}(\vv x; \vv\vartheta)$:
\begin{align}
\label{eq:af_inverse_transform}
z_t =  x_t \cdot \sigma_t (x_{<t}; \vv\vartheta) + 
\mu_t (x_{<t}; \vv\vartheta),
\end{align}
where the shifting variables $\mu_t (x_{<t}; \vv\vartheta)$ and scaling variables $\sigma_t (x_{<t}; \vv\vartheta)$ are modeled by an autoregressive architecture parameterized by $\vv\vartheta$~(e.g., WaveNet).
Note that, the $t$-th variable $z_t$ only depends on $x_{\leq t}$, thus the Jacobian is a triangular matrix as illustrated in Figure~\ref{fig:Jacobian}(a). Its determinant is the product of the diagonal entries: $\det\left( \frac{\partial f^{-1}(\vv x) }{\partial \vv x} \right) = \prod_{t} \sigma_t (x_{<t}; \vv\vartheta)$.
The density $p(\vv x)$ can be evaluated in parallel by  formula~Eq.~\eqref{eq:change_of_variable}, because the minimum number of sequential operations is $O(1)$ for computing $\vv z = f^{-1}(\vv x)$~(see Table~\ref{tab:summary_flow}). 
However, AF has to do sequential synthesis, because  $\vv x = f(\vv z)$ is autoregressive: 
$
x_t =  \frac{z_t - \mu_t (x_{<t}; \vv\vartheta)}{\sigma_t (x_{<t}; \vv\vartheta)}.
$
Note that, Gaussian autoregressive model can be equivalently interpreted as an autoregressive flow~\citep{kingma2016improved}.

In contrast, IAF uses an autoregressive transformation for inverse mapping $\vv z = f^{-1}(\vv x)$:
\begin{align}
\label{eq:iaf_inverse_transform}
z_t =  \frac{x_t - \mu_t (z_{<t}; \vv\vartheta)}{\sigma_t (z_{<t}; \vv\vartheta)},
\end{align}
making density evaluation very slow for likelihood-based training, but one can sample $\vv x = f(\vv z)$ in parallel through
$x_t =  z_t \cdot \sigma_t (z_{<t}; \vv\vartheta) + 
\mu_t (z_{<t}; \vv\vartheta)$.
Parallel WaveNet \citep{oord2017parallel} and ClariNet~\citep{ping2018clarinet} are based on IAF for parallel synthesis, and they rely on the probability density distillation from a pretrained autoregressive WaveNet at training.

\begin{table*}[t!]
\centering
\vspace{-.3em}
\caption{The minimum number of sequential operations~(indicates parallelism) required by flow-based models for density evaluation $\vv z = f^{-1}(\vv x)$ and sampling $\vv x = f(\vv z)$. Therein, $n$ is the length of $\vv x$, $h$ is the squeezed height in WaveFlow. In WaveFlow, larger $h$ leads to higher model capacity, but more sequential steps for sampling.}
\vspace{0.2em}
\begin{tabular}{l|c|c|c}
\hline 
\multirow{2}{8.5em}{Flow-based model}  &  Sequential operations & Sequential operations & Model capacity \\
& for $\vv z = f^{-1}(\vv x)$ & for $\vv x = f(\vv z)$ & (same size)
\\ \hline
AF & $O(1)$ & $O(n)$ & high  \\ 
IAF  & $O(n)$ & $O(1)$ & high \\ 
Bipartite flow  & $O(1)$ & $O(1)$ & low  \\
WaveFlow & $O(1)$ & $O(h)$ & low $\leftrightarrow$ high \\ 
\hline
\end{tabular}
\label{tab:summary_flow}
\end{table*}

\vspace{-.2em}
\subsection{Bipartite transformation}
\vspace{-.1em}
RealNVP~\citep{dinh2016density} and Glow~\citep{kingma2018glow} use bipartite transformation by partitioning the data $\vv x$ into two groups $\vv x_a$ and $\vv x_b$, where the indices sets $a\cup b = \{1, \cdots, n \}$ and $a\cap b = \phi$. Then, the inverse mapping $\vv z = f^{-1}(\vv x, \vv\theta)$ is defined as:
\begin{align}
    z_a = x_a, \quad
    z_b = x_b \cdot \sigma_b(x_a; \vv\theta) + \mu_b(x_a; \vv\theta).
\end{align}
where the shifting variables $\mu_b(x_a; \vv\theta)$ and scaling variables $\sigma_b(x_a; \vv\theta)$ are modeled by a feed-forward neural network.
Its Jacobian $\frac{\partial f^{-1}(\vv x) }{\partial \vv x}$ is a special triangular matrix as illustrated in Figure~\ref{fig:Jacobian}(b).
By definition,  $\vv x = f(\vv z, \vv\theta)$ is,
\begin{align}
    x_a = z_a, \quad
    x_b = \frac{z_b - \mu_b(x_a; \vv\theta)}{\sigma_b( x_a; \vv\theta)}.
\end{align}
Note that, both evaluating $\vv z = f^{-1}(\vv x, \vv\theta)$ and sampling  $\vv x = f(\vv z, \vv\theta)$ can be done in parallel.

WaveGlow~\citep{prenger2019waveglow} and FloWaveNet~\citep{kim2018flowavenet} \emph{squeeze} the time-domain samples on the \emph{channel} dimension, then apply the bipartite transformation  on the partitioned channels. 
Note that, this squeezing operation is inefficient, as one may lose the temporal order information.
As a result of doing this, the synthesized audio can have constant frequency noises~(see Appendix~A for an example).

\vspace{-.2em}
\subsection{Connections}
\label{subsec:auto_vs_bipartite}
\vspace{-.1em}
The autoregressive transformation is more expressive than bipartite transformation.
As illustrated in Figure~\ref{fig:Jacobian}(a) and~(b), the autoregressive transformation introduces $\frac{n\times(n-1)}{2}$ complex non-linear dependencies~(dark-blue cells) and $n$ linear dependencies between data $\vv x$ and latents $\vv z$. 
In contrast, bipartite transformation has only $\frac{n^2}{4}$ non-linear dependencies and $\frac{n}{2}$ linear dependencies.
Indeed, one can easily reduce an autoregressive transformation $\vv z = f^{-1}(\vv x; \vv\vartheta)$ to a bipartite transformation $\vv z = f^{-1}(\vv x; \vv\theta)$ by: (\emph{i})~picking an autoregressive order $\vv o$ such that all of the indices in set $a$ rank early than the indices in $b$, and (\emph{ii}) setting the shifting and scaling variables as,
\vspace{-.6em}
\begin{align*}
\begin{pmatrix}
\mu_{t}(x_{<t}; \vv\vartheta) \\
\sigma_{t}(x_{<t}; \vv\vartheta)
\end{pmatrix}
 = 
\begin{cases}
\big(0,~ 1\big)^{\top},   &\text{for $t\in a$}\\
\big( \mu_{t}(x_{a}; \vv\theta),~ \sigma_{t}(x_{a}; \vv\theta) \big)^\top, & \text{for $t \in b$}
\end{cases}.
\vspace{-1.5em}
\end{align*}
Given the less expressive building block, the bipartite flows require more layers and larger hidden size to reach the capacity of autoregressive models~(e.g., measured by likelihood).

\vspace{-.2em}
\section{WaveFlow}
\label{sec:waveflow}
\vspace{-.1em}
In this section, we present WaveFlow and its implementation with dilated 2-D convolutions. We also discuss the permutation strategies for stacking multiple flows.
\vspace{-.2em}
\subsection{Definition}
\label{subsec:definition}
\vspace{-.1em}
\begin{figure*}[t] \centering
\begin{tabular}{ccc}
\includegraphics[height=1.9cm, clip]{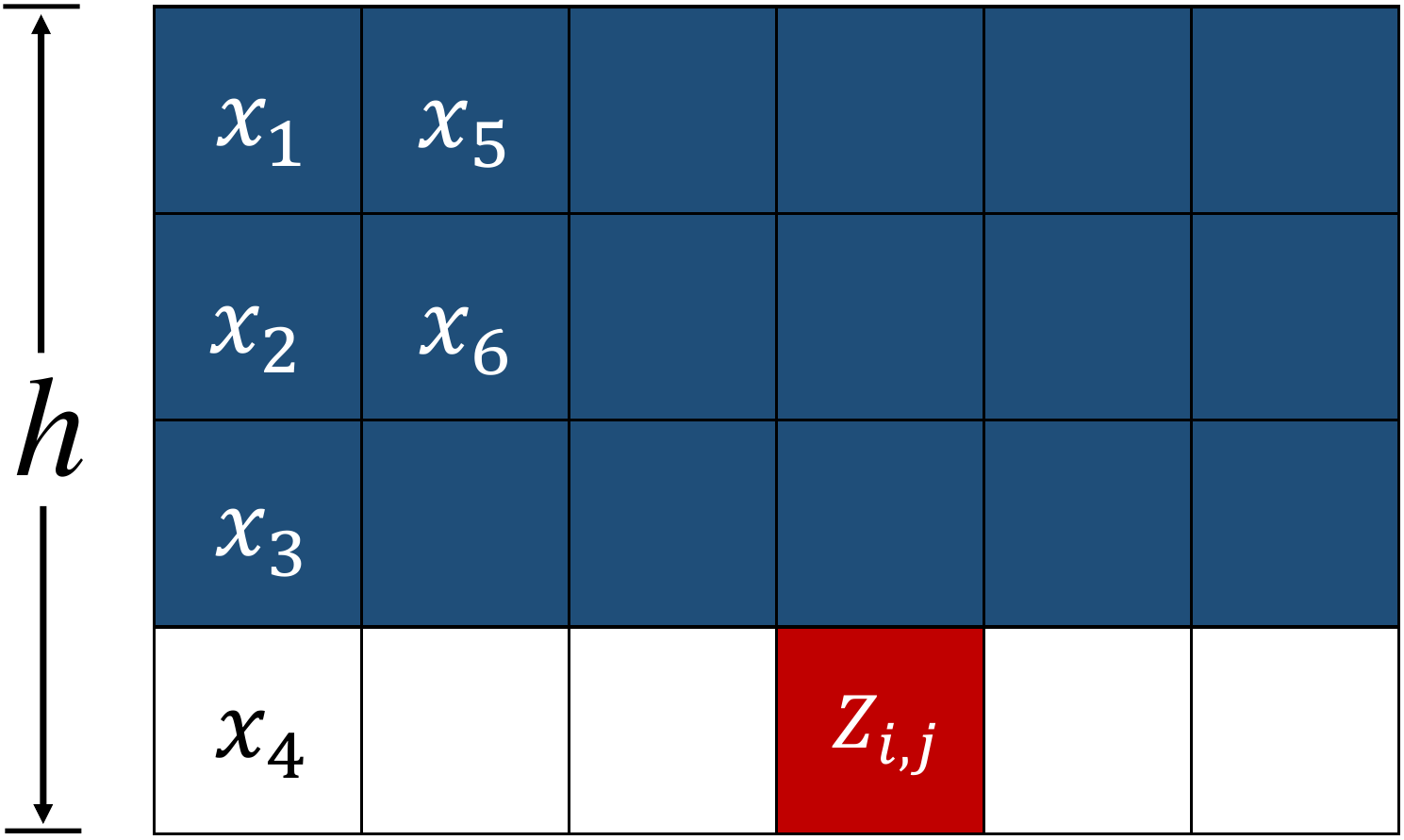} 
&\hspace{1.0cm}
\includegraphics[height=1.9cm, clip]{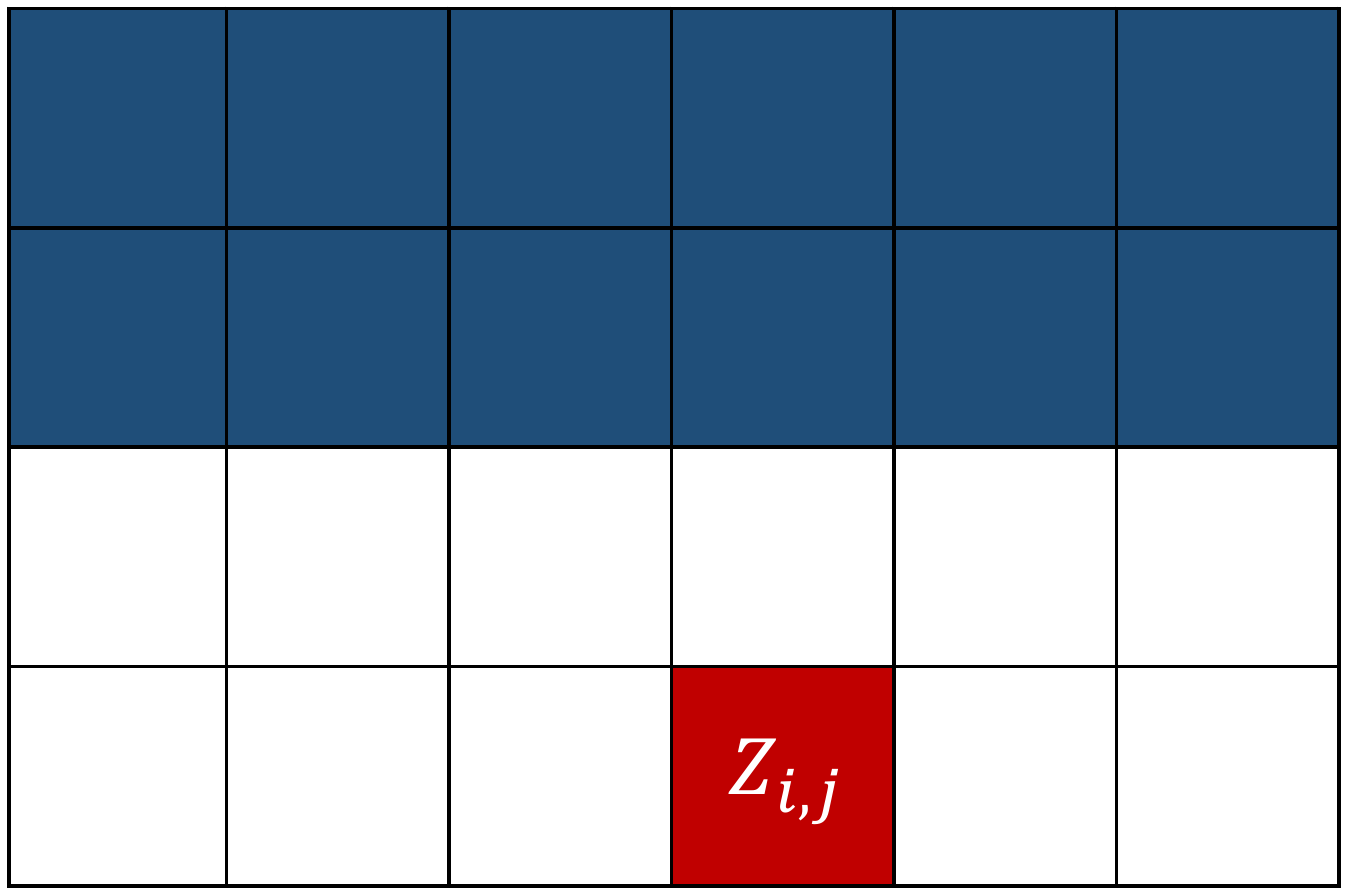} 
&\hspace{1.0cm}
\includegraphics[height=1.9cm, clip]{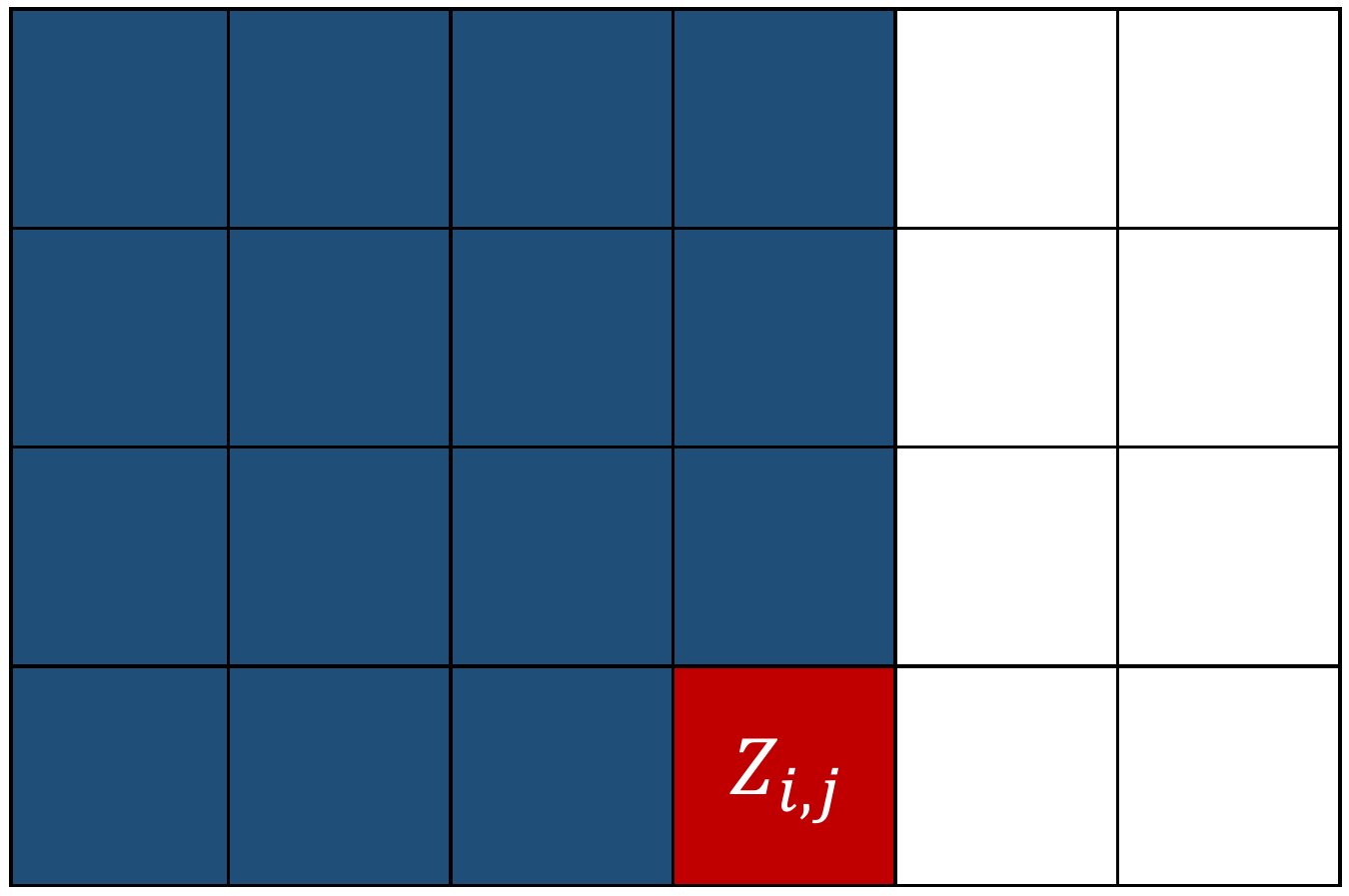} 
\\
\vspace{-.3em}
{\small (a)}  & \hspace{1.0cm}  {\small(b)} & \hspace{1.0cm} {\small(c)} 
\\
\end{tabular}
\vspace{-.5em}
\caption{The receptive fields over the squeezed inputs $X$ for computing $Z_{i,j}$ in  (a)~WaveFlow, (b)~WaveGlow, and (c)~autoregressive flow with column-major order~(e.g., WaveNet).}
\vspace{-.3em}
\label{fig:Receiptive_field} %
\end{figure*}
We denote the 1-D waveform as $\vv x = \{x_1, \cdots, x_n \}$.
We \emph{squeeze} $\vv x$ into an $h$-row 2-D matrix $X \in \mathbb{R}^{h \times w}$ by column-major order, where the adjacent samples are in the same column. 
We assume $Z \in \mathbb{R}^{h \times w}$ are sampled from an isotropic Gaussian distribution, and define $Z = f^{-1}(X; \Theta)$ as,
\begin{align}
\label{eq:waveflow_inverse_transform}
Z_{i, j} = \sigma_{i,j} (X_{<i, \bullet}; \Theta) \cdot X_{i,j} + \mu_{i,j}(X_{<i, \bullet}; \Theta),
\end{align}
where $X_{<i, \bullet}$ represents all elements above $i$-th row~(see Figure~\ref{fig:Receiptive_field} for an illustration).
Note that, (\emph{i}) In WaveFlow, the receptive field over the squeezed inputs $X$ for computing $Z_{i,j}$  is strictly larger than that of WaveGlow when $h>2$.
(\emph{ii})~WaveNet is equivalent to an autoregressive flow~(AF) with the column-major order on $X$. 
(\emph{iii})~Both WaveFlow and WaveGlow look at future waveform samples in original $\vv x$ for computing $Z_{i, j}$, whereas WaveNet can not.

The shifting variables $\mu_{i,j}(X_{<i, \bullet}; \Theta)$ and scaling variables $\sigma_{i,j} (X_{<i, \bullet}; \Theta)$ in Eq.~\eqref{eq:waveflow_inverse_transform} are modeled by a 2-D convolutional neural network detailed in Section~\ref{subsec:conv2d}.
By definition, the variable $Z_{i, j}$ only depends on the current $X_{i, j}$ and previous $X_{<i, \bullet}$ in raw-major order, thus the Jacobian is a triangular matrix and its determinant is:
\begin{align}
\det\left( \frac{\partial f^{-1}(X) }{\partial X} \right) = \prod_{i=1}^{h}\prod_{j=1}^{w} \sigma_{i, j} (X_{<i, \bullet}; \Theta).
\end{align}
As a result, the log-likelihood can be calculated in parallel by change of variable formula in  Eq.~\eqref{eq:change_of_variable},
\begin{align}
\log p(X) = \sum_{i, j} \Big( \log \sigma_{i, j} (X_{<i, \bullet}; \Theta) -   \frac{Z_{i, j}^2}{2} - \frac{1}{2}\log(2\pi) \Big), \nonumber
\end{align}
and one can do maximum likelihood training efficiently.
At synthesis, we sample $Z$ from the isotropic Gaussian and apply the forward mapping $X = f(Z; \Theta)$:
\begin{align}
\label{eq:waveflow_forward}
X_{i,j} = \frac{Z_{i, j} - \mu_{i,j}(X_{<i, \bullet}; \Theta)} {\sigma_{i,j} (X_{<i, \bullet}; \Theta)},
\end{align}
which is autoregressive over the height dimension and requires $h$ sequential steps to generate the whole $X$. In practice, a small $h$ (e.g., $8$ or $16$) works well, thus we can generate very long waveforms within a few sequential steps.

\vspace{-.2em}
\subsection{Implementation with dilated 2-D convolutions}
\label{subsec:conv2d}
\vspace{-.1em}
We implement WaveFlow with a dilated 2-D convolutional architecture.
Specifically, we use a stack of 2-D convolution layers~(e.g., 8 layers in all our experiments) to model the shifting $\mu_{i,j}(X_{<i, \bullet}; \Theta)$ and scaling variables $\sigma_{i,j} (X_{<i, \bullet}; \Theta)$ in Eq.~\eqref{eq:waveflow_inverse_transform}. 
We use the similar architecture as WaveNet~\citep{oord2016wavenet} by replacing the dilated 1-D convolution with 2-D convolution~\citep{yu2015multi}, while still keeping the gated-tanh nonlinearities, residual connections and skip connections.

We set the filter sizes as 3 for both height and width dimensions.
We use non-causal convolutions on width dimension and set the dilation cycle as $[1, 2, 4, \cdots, 2^7]$.
The convolutions on height dimension are causal with the autoregressive constraint, 
and their dilation cycle needs to be designed carefully. In practice, we find the following rules of thumb are important to obtain good results:
\begin{itemize}[noitemsep,topsep=0pt, leftmargin=1.75em]
    \vspace{-0.1cm}
    \item As motivated by the dilation cycle of WaveNet~\citep{oord2016wavenet}, the dilations of 8 layers should be set as $\vv d = [1, 2, \cdots, 2^s, 1, 2, \cdots, 2^s, \cdots]$, where $s\le7$.~\footnote{We did try different setups, but they all lead to worse likelihood scores.}
    \item The receptive field $r$ over the height dimension should be larger than or equal to height~$h$. Otherwise, it introduces unnecessary conditional independence and leads to lower likelihood (see Table~\ref{tab:small_receptive_field_poor_ll} for an example). 
    Note that, the receptive field of a stack of dilated convolutional layers is: $r = (k - 1) \times \sum_i d_i + 1$, where $k$ is the filter size and $d_i$ is the dilation at $i$-th layer. Thus, the sum of dilations should satisfy: $\sum_i d_i \ge \frac{h - 1}{k - 1}$.
    However, when $h$ is larger than or equal to $2^8 = 512$, we simply set the dilation cycle as $[1, 2, 4, \cdots, 2^7]$.
    \item When $r$ has already been larger than $h$, the convolutions with smaller dilations provide larger likelihood.
    \vspace{-.2em}
\end{itemize}

\begin{table*}[t!]
\centering
\caption{The test log-likelihoods~(LLs) of WaveFlow with different dilation cycles on the height dimension when $h = 32$. The models are stacked with $8$ flows and each flow has $8$ layers.}
\vspace{0.2em}
\begin{tabular}{l|c|c|c|c}
\hline 
Model & Res. channels  &  Dilations $\vv d$ & Receptive field $r$ & Test LLs
\\ \hline
WaveFlow~($h = 32$) & 128 & $ 1, 1, 1, 1, 1, 1, 1, 1$ & 17 & $4.960$ \\ 
WaveFlow~($h = 32$) & 128 & $ 1, 2, 4, 1, 2, 4, 1, 2$ & 35 & $5.055$ \\ 
\hline
\end{tabular}
\label{tab:small_receptive_field_poor_ll}
\end{table*}
\begin{table}[t!]
\centering
\vspace{-0.4cm}
\caption{The height $h$, filter size $k$ over the height dimension, and the corresponding dilations used in our experiments. Note that, the receptive fields $r$ are only slightly larger than heights $h$.}
\vspace{0.2em}
\begin{tabular}{l|c|l|c}
\hline 
 $h$ & $k$  & \quad Dilations $\vv d$   & Receptive field $r$
\\ \hline
8  & 3 & $ 1, 1, 1, 1, 1, 1, 1, 1$ & 17 \\ 
16 & 3 & $ 1, 1, 1, 1, 1, 1, 1, 1$ & 17 \\ 
32 & 3 & $ 1, 2, 4, 1, 2, 4, 1, 2$ & 35 \\ 
64 & 3 & $ 1, 2, 4, 8, 16, 1, 2, 4$ & 77 \\ 
\hline
\end{tabular}
\vspace{-0.15cm}
\label{tab:dilations}
\end{table}

We summarize the heights and preferred dilations in our experiments in Table~\ref{tab:dilations}.
We also implement \emph{convolution queue}~\citep{paine2016fast} to cache the intermediate hidden states, which will speed up the autoregressive inference over the height dimension.
Note that, WaveFlow is fully autoregressive and equivalent to a Gaussian WaveNet~\citep{ping2018clarinet}, when we \emph{squeeze} $\vv x$ by its length~(i.e. $h = n$) and set its filter size as $1$ over the width dimension. 
If we squeeze $\vv x$ by $h=2$ and set the filter size as $1$ on height dimension, WaveFlow becomes a bipartite flow and is equivalent to a WaveGlow with squeezed channels 2.

\vspace{-.2em}
\subsection{Local conditioning for speech synthesis}
\vspace{-.1em}
In neural speech synthesis, a neural vocoder~(e.g., WaveNet) synthesizes the time-domain waveforms, which can be conditioned on linguistic features~\citep{oord2016wavenet,arik2017DV1}, the mel spectrograms from a text-to-spectrogram model~\citep{ping2017deep, shen2018tacotron2}, or the learned hidden representation within a text-to-wave architecture~\citep{ping2018clarinet}. 
In this work, we test WaveFlow by conditioning it on ground-truth mel spectrograms as in previous work~\citep{prenger2019waveglow,kim2018flowavenet}.
The mel spectrogram is upsampled to the same length as waveform samples with transposed 2-D convolutions~\citep{ping2018clarinet}.
To be aligned with the waveform, they are squeezed to the shape $c\times h\times w$, where $c$ is the input channels~(e.g, mel bands).
After a $1\times1$ convolution mapping the input channels to residual channels, they are added as the bias term at each layer~\citep{ping2018clarinet}.

\vspace{-.1em}
\subsection{Stacking multiple flows with permutations on height dimension}
\label{subsec:stacking_permutate}
\vspace{-0.1em}
\begin{table*}[t!]
\centering
\caption{The test LLs of WaveFlow with different permutation strategies. All models consist of $8$ flows and each flow has $8$ convolutional layers with filter sizes $3$.}
\vspace{0.1em}
\begin{tabular}{l|c|l|c}
\hline 
{Model} & Res. channels  & \qquad\qquad Permutation strategy & Test LLs
\\ \hline
WaveFlow~($h=16$) & 64 &  none & $4.551$ \\ 
WaveFlow~($h=16$) & 64 & \emph{a}) 8 reverse   & $4.954$ \\ 
WaveFlow~($h=16$) & 64 & \emph{b}) 4 reverse, 4 bipartition \& reverse & $4.971$ \\ 
\hline
\end{tabular}
\label{tab:lls_reverse}
\end{table*}
Flow-based models require a series of transformations until the distribution $p(X)$ reaches a desired level of capacity.
We denote $X = Z^{(n)}$ and repeatedly apply the transformation $Z^{(i-1)} = f^{-1} (Z^{(i)}; \Theta^{(i)})$ defined in Eq.~\eqref{eq:waveflow_inverse_transform} from $Z^{(n)}$ to  $Z^{(0)}$, where $Z^{(0)}$ are from the isotropic Gaussian.
Thus, the $p(X)$ can be evaluated by applying the chain rule:
$$
p(X) = p(Z^{(0)}) \prod_{i=1}^n \left| \det\left( \frac{\partial f^{-1}(Z^{(i)}; \Theta^{(i)}) }{\partial Z^{(i)}} \right) \right| .
$$
We find that permuting each $Z^{(i)}$ over its height dimension after each transformation can significantly improve the likelihood scores.
In particular, we test two permutation strategies for WaveFlow models stacked with 8 flows~(i.e., $X = Z^{(8)}$) in Table~\ref{tab:lls_reverse}: \emph{a}) we reverse each $Z^{(i)}$ over the height dimension after each transformation, and \emph{b}) we reverse $Z^{(7)}$, $Z^{(6)}$, $Z^{(5)}$, $Z^{(4)}$ over the height dimension as before, but bipartition $Z^{(3)}$, $Z^{(2)}$, $Z^{(1)}$, $Z^{(0)}$ in the middle of the height dimension then reverse each part respectively.~\footnote{After bipartition \& reverse, the height dimension [0, $\cdots$, $\frac{h}{2}-1$, $\frac{h}{2}$, $\cdots$ , $h-1$] becomes [$\frac{h}{2}-1$, $\cdots$, 0, $h-1$, $\cdots$, $\frac{h}{2}$].}
In speech synthesis, one needs to permute the conditioner accordingly over the height dimension, which is aligned with $Z^{(i)}$.
In Table~\ref{tab:lls_reverse}, both strategies \emph{a}) and \emph{b}) significantly outperform the model without permutations mainly because of bidirectional modeling. Strategy~\emph{b}) outperforms \emph{a}), and we attribute this to its diverse autoregressive orders.

\vspace{-.3em}
\section{Related work}
\label{sec:related_work}
\vspace{-.1em}
Neural speech synthesis has obtained the state-of-the-art results and received a lot of attention. 
Several neural text-to-speech~(TTS) systems have been introduced, including WaveNet~\citep{oord2016wavenet}, Deep Voice~1~\&~2~\&~3 \citep{arik2017DV1, arik2017DV2, ping2017deep}, Tacotron 1~\&~2 \citep{wang2017tacotron, shen2018tacotron2}, Char2Wav \citep{sotelo2017char2wav}, VoiceLoop~\citep{taigman2018voiceloop}, WaveRNN~\citep{kalchbrenner2018efficient}, ClariNet~\citep{ping2018clarinet}, Transformer TTS~\citep{li2019neural}, ParaNet~\citep{peng2019parallel}, and FastSpeech~\citep{ren2019fastspeech}.

Neural vocoders~(waveform synthesizer), such as WaveNet,  play the most important role in recent advances of speech synthesis. 
In previous work, the state-of-the-art neural vocoders are autoregressive models~\citep{oord2016wavenet, mehri2016samplernn, kalchbrenner2018efficient}.
Several endeavors have been advocated for speeding up their sequential generation process~\citep{arik2017DV1, kalchbrenner2018efficient}.
In particular, Subscale WaveRNN~\citep{kalchbrenner2018efficient} folds a long waveform sequence $\vv x_{1:n}$ into a batch of shorter sequences and can produces up to $16$ samples per step, thus it requires at least $\frac{n}{16}$ steps to generate the whole audio.
This is different from WaveFlow, which can generate $\vv x_{1:n}$ within e.g. 16 steps.

Flow-based models can either represent the approximate posteriors for variational inference~\citep{rezende2015variational, kingma2016improved, berg2018sylvester}, or can be trained directly on data using the change of variables formula~\citep{dinh2014nice,dinh2016density,kingma2018glow}.
We consider the later case in this work.
In previous work, Glow~\citep{kingma2018glow} extends RealNVP~\citep{dinh2016density} with invertible $1\times1$ convolution on channel dimension, which first generates high-fidelity images. 
\citet{hoogeboom2019emerging} generalizes the  invertible convolution to operate on both channels and spatial axes.
Most recently, flow-based models have been successfully applied for parallel waveform synthesis with comparable fidelity as autoregressive models~\citep{oord2017parallel, ping2018clarinet, yamamoto2019probability, prenger2019waveglow, kim2018flowavenet, serra2019blow}.
Among these models, WaveGlow~\citep{prenger2019waveglow} and FloWaveNet~\citep{kim2018flowavenet} have a simple training pipeline as they solely use the maximum likelihood objective.
However, both of them are less expressive than autoregressive models as indicated by their large memory footprint and lower likelihood scores.

\section{Experiment}
\label{sec:experiment}
%
In this section, we compare likelihood-based generative models for raw audio in term of test likelihood, synthesis speed and speech fidelity.~\footnote{\small{Speech samples are in:~\url{https://waveflow-demo.github.io/}}}
The results in this section are obtained from an internal PyTorch implementation. We provide a PaddlePaddle reimplementation in Parakeet toolkit.~\footnote{\small{\url{https://github.com/PaddlePaddle/Parakeet/tree/develop/examples/waveflow}}}

\vspace{-.3em}
{\bf Data:} We use the LJ speech dataset~\citep{Ito2017ljspeech} containing about 24 hours of audio with a sampling rate of 22.05~kHz recorded on a MacBook Pro in a home enviroment.
It consists of $13,100$ audio clips from a single female speaker.

\vspace{-.3em}
{\bf Models:}
We evaluate several likelihood-based models, including WaveFlow,  Gaussian WaveNet~\citep{ping2018clarinet}, WaveGlow~\citep{prenger2019waveglow}, and autoregressive flow~(AF). 
As illustrated in Section~\ref{subsec:conv2d}, we implement AF from WaveFlow by squeezing the waveforms by its length and setting the filter size as 1 over width dimension.
Both WaveNet and AF have 30 layers with dilation cycle $[1, 2, \cdots, 512]$ and filter size 3.
For WaveFlow and WaveGlow, we investigate different setups, including the number of flows, size of residual channels, and squeezed height $h$.

\vspace{-.3em}
{\bf Conditioner:}
We use the 80-band mel spectrogram of the original audio as the conditioner for WaveNet, WaveGlow, and WaveFlow.
We set FFT size to 1024, hop size to 256, and window size to 1024.
For WaveNet and WaveFlow, we upsample the mel conditioner 256 times by applying two layers of transposed 2-D convolution~(in time and frequency) interleaved with leaky ReLU~($\alpha=0.4$)~\citep{ping2018clarinet}.
The upsampling strides in time are $16$ and the 2-D convolution filter sizes are $[32, 3]$ for both layers.
In WaveFlow, the upsampled mel spectrogram is squeezed along the temporal dimension as waveform and its shape becomes [mel-band, height, width]. After that, we apply 1x1 conv to map its channels from mel-band to the residual channel in each 2-D convolution layer. Finally, it is added as bias term within the dilated convolution operation before the gated-tanh nonlinearities, which is the same as WaveNet
For WaveGlow, we directly use Nvidia's open source implementation.

\vspace{-.3em}
{\bf Training:} 
We train all models on 8 Nvidia 1080Ti GPUs using
randomly chosen short clips of $16,000$ samples from each utterance.
For WaveFlow and WaveNet, we use the Adam optimizer~\citep{kingma2014adam} with a batch size of~8 and a constant learning rate of $2\times10^{-4}$.
For WaveGlow, we use the Adam optimizer with a batch size of 16 and a learning rate of $1\times10^{-4}$.
We apply weight normalization~\citep{salimans2016weight}  whenever possible.

\vspace{-.2em}
\subsection{Likelihood}
\vspace{-.1em}
\begin{table*}[t]
\centering
\caption{The test log-likelihoods~(LLs) of all models conditioned on mel spectrograms.
For $a\times b = c$ in the {"flows$\times$layers"} column, $a$ is number of flows, $b$ is number of layers in each flow, and $c$ is the total number of layers.
In WaveFlow, $h$ is the squeezed height. Models with bolded test LLs are mentioned in the text.}
\vspace{0.25em}
\begin{tabular}{l|l|r|r|r|c}
\hline 
& \textbf{Model} & \textbf{flows$\times$layers} & \textbf{Res. channels} &  \textbf{\# Param} 
& \textbf{Test LLs}   \\ \hline
(\emph{a}) & Gaussian WaveNet   & 1$\times$30 = 30  \qquad & 128 \qquad & 4.57 M \qquad &  $\bf{5.059}$  \\
(\emph{b}) & Autoregressive flow  & 3$\times$10 = 30  \qquad & 128 \qquad & 4.54 M \qquad &  $\bf{5.161}$  \\
(\emph{c}) & WaveGlow   & 12$\times$8 = 96 \qquad & 64  \qquad & 17.59 M \qquad  &   $4.804$  \\
(\emph{d}) & WaveGlow  & 12$\times$8 = 96  \qquad & 128 \qquad & 34.83 M \qquad &   $4.927$  \\
(\emph{e}) & WaveGlow  & 6$\times$8 = 48  \qquad & 256 \qquad &  47.22 M \qquad &   $4.922$  \\
(\emph{f}) & WaveGlow  & 12$\times$8 = 96  \qquad & 256 \qquad & 87.88 M \qquad &   $5.018$  \\
(\emph{g}) & WaveGlow  & 12$\times$8 = 96  \qquad & 512 \qquad & 268.29 M \qquad &  $\bf{5.026}$ \\ 
(\emph{h}) & WaveFlow ($h = 8$)  & 8$\times$8 = 64  \qquad & 64 \qquad & 5.91 M \qquad &  $4.935$  \\
(\emph{i}) & WaveFlow ($h = 16$) & 8$\times$8 = 64  \qquad & 64 \qquad & 5.91 M \qquad &  $4.954$  \\
(\emph{j}) & WaveFlow~($h = 32$) & 8$\times$8 = 64  \qquad & 64 \qquad & 5.91 M \qquad &  $5.002$  \\
(\emph{k}) & WaveFlow~($h = 64$) & 8$\times$8 = 64  \qquad & 64 \qquad & 5.91 M \qquad &  $\bf{5.023}$   \\
(\emph{l}) & WaveFlow~($h = 8$)  & 6$\times$8 = 48  \qquad & 96 \qquad & 9.58 M \qquad &  $4.946$  \\
(\emph{m}) & WaveFlow~($h = 8$)  & 8$\times$8 = 64  \qquad & 96 \qquad & 12.78 M \qquad &  $4.977$  \\
(\emph{n}) & WaveFlow~($h = 16$)  & 8$\times$8 = 64  \qquad & 96 \qquad & 12.78 M \qquad &  $5.007$  \\
(\emph{o}) & WaveFlow~($h = 16$)     & 6$\times$8 = 48  \qquad & 128 \qquad &  16.69 M \qquad & $4.990$ \\ 
(\emph{p}) & WaveFlow~($h = 8$)     & 8$\times$8 = 64  \qquad & 128 \qquad & 22.25 M \qquad & $5.009$ \\ 
(\emph{q}) & WaveFlow~($h = 16$)    & 8$\times$8 = 64  \qquad & 128 \qquad & 22.25 M \qquad & $5.028$ \\ 
(\emph{r}) & WaveFlow~($h = 32$)    & 8$\times$8 = 64  \qquad & 128 \qquad & 22.25 M \qquad & $\bf{5.055}$ \\ 
(\emph{s}) & WaveFlow~($h = 16$)    & 6$\times$8 = 48  \qquad & 256 \qquad & 64.64 M \qquad & $5.064$  \\
(\emph{t}) & WaveFlow~($h = 16$)    & 8$\times$8 = 64  \qquad & 256 \qquad & 86.18 M  \qquad & $\bf{5.101}$  \\ \hline
\end{tabular}
\label{tab:test_likelihood}
\vspace{-.2em}
\end{table*}
We evaluate the test log-likelihoods~(LLs) of WaveFlow, WaveNet, WaveGlow  and autoregressive flow~(AF) conditioned on mel spectrograms at 1M training steps. 
We choose 1M steps as the cut-off, because the LLs decrease slowly after that, and it already took one month to train the largest WaveGlow~(residual channels = 512) for 1M steps.
We summarize the results in Table~\ref{tab:test_likelihood} with models from row (\emph{a}) to~(\emph{t}).
We draw the following observations:
\begin{itemize}[itemsep=0.00pt, topsep=0pt, leftmargin=1.25em]
    \vspace{-.2em}
    \item Stacking a large number of flows improves LLs for all flow-based models.
    For example, WaveFlow~(\emph{m}) with 8 flows provides larger LL than WaveFlow~(\emph{l}) with 6 flows.
    The autoregressive flow~(\emph{b}) obtains the highest likelihood and outperforms WaveNet~(\emph{a}) with the same amount of parameters. Indeed, AF provides bidirectional modeling by stacking 3 flows with reverse operations.
    \vspace{-.2em}
    \item WaveFlow has much larger likelihood than WaveGlow with comparable number of parameters.
     In particular, a small-footprint WaveFlow~(\emph{k}) has only 5.91M parameters but can provide comparable likelihood~(5.023 vs. 5.026) as the largest WaveGlow~(\emph{g}) with 268.29M parameters.
    \vspace{-.2em}
    \item As we increase $h$, the likelihood of WaveFlow steadily increases~(can be seen from (\emph{h})-(\emph{k})), and its inference will be slower on GPU with more sequential steps. In the limit, it is equivalent to an AF. It illustrates the trade-off between model capacity and inference parallelism.
    \vspace{-.2em}
    \item WaveFlow~(\emph{r}) with 128 residual channels can obtain comparable likelihood~(5.055 vs 5.059)  as WaveNet~(\emph{a}) with 128 residual channels.
    A larger WaveFlow~(\emph{t}) with 256 residual channels can obtain even larger likelihood than WaveNet~(5.101 vs 5.059).
    \vspace{-.1em}
\end{itemize}
Note that, there is a significant likelihood gap that has so far existed between autoregressive models and flow-based models providing efficient sampling~\citep[e.g.,][]{ho2019flow++, tran2019discrete}. WaveFlow can close the likelihood gap with a modest squeezing height $h$, which suggests the strength of autoregressive model is mainly at modeling the local structure of the signal.

\vspace{-.2em}
\subsection{Speech fidelity and synthesis speed}
\vspace{-.1em}
We use the permutation strategy \emph{b}) described in Table~\ref{tab:lls_reverse} for WaveFlow.
We train WaveNet for 1M steps.
We train large WaveGlow and WaveFlow~(res. channels 256 and 512) for 1M steps due to the practical time constraint.
We train moderate size models~(res. channels 128) for 2M steps.
We train small size models~(res. channels 64 and 96) for 3M steps with slightly improved performance after 2M steps.
We use the same setting of ClariNet as in~\citet{ping2018clarinet}.
At synthesis, we sample $Z$ from an isotropic Gaussian with standard deviation 1.0 and 0.6~(default) for WaveFlow and WaveGlow, respectively.
We use the crowdMOS tookit~\citep{ribeiro2011crowdmos} for speech quality evaluation, where test utterances from these models were presented to workers on Mechanical Turk. 
In addition, we test the synthesis speed on a Nvidia V100 GPU without using any engineered inference kernels. 
For WaveFlow and WaveGlow, we run synthesis under NVIDIA Apex with 16-bit floating point (FP16) arithmetic, which does not introduce any degradation of audio fidelity and brings about $2\times$ speedup. 
We implement \emph{convolution queue}~\citep{paine2016fast} in Python to cache the intermediate hidden states in WaveFlow for autoregressive inference over the height dimension, which brings another  $3\times$ to $5\times$ speedup depending on height $h$. 
\begin{table*}[t!]
\centering
\caption{The model size, synthesis speed over real-time, and the 5-scale Mean Opinion Scores (MOS) with 95$\%$ confidence intervals. 
}
\vspace{0.2em}
\begin{tabular}{l|r|r|r|r|c}
\hline 
\textbf{Model}  & \textbf{flows$\times$layers} & \textbf{res. channels} & \textbf{\# param} & \textbf{syn. speed} & \textbf{MOS}
\\ \hline
Gaussian WaveNet & 1$\times$30 = 30  & 128 \qquad &  {4.57} M & ${0.002\times}$\qquad & ${\bf 4.43 \pm 0.14}$   \\ 
ClariNet & 6$\times$10 = 60  & 64 \qquad &  2.17 M & ${21.64\times}$\qquad & ${4.22 \pm 0.15}$   \\ 
WaveGlow & 12$\times$8 = 96   & {64} \qquad & 17.59 M & $93.53\times$\qquad & ${2.17 \pm 0.13}$   \\ 
WaveGlow & 12$\times$8 = 96  & 128 \qquad & 34.83 M & $69.88\times$\qquad & $2.97 \pm 0.15$      \\ 
WaveGlow & 12$\times$8 = 96  & {256} \qquad & {\bf 87.88} M & ${\bf 34.69\times}$\qquad & ${\bf 4.34 \pm 0.11}$   \\ 
WaveGlow & 12$\times$8 = 96  & 512 \qquad & 268.29 M & $8.08\times$\qquad  & $4.32 \pm 0.12$  \\ 
WaveFlow~($h=8$) & 8$\times$8 = 64 & 64 \qquad & 5.91 M &  $47.61\times$ \qquad & $4.26 \pm 0.12$   \\ 
WaveFlow~($h=16$) & 8$\times$8 = 64 & {64} \qquad & {\bf 5.91} M &  ${\bf 42.60\times}$
\qquad & ${\bf 4.32 \pm 0.08}$   \\ 
WaveFlow~($h=16$) & 8$\times$8 = 64 &  96 \qquad & 12.78 M & $26.23\times$ \qquad & $4.34 \pm 0.13$   \\ 
WaveFlow~($h=16$) & 8$\times$8 = 64  & 128 \qquad & 22.25 M &  $21.32\times$
\qquad & $4.38 \pm 0.09$  \\ 
WaveFlow~($h=16$) & 8$\times$8 = 64 & {256} \qquad  & {86.18} M  &  ${8.42\times}$
\qquad  & ${\bf 4.43 \pm 0.10}$   \\ \hline
Ground-truth & \NA \qquad & \NA\qquad & \NA \qquad & \NA \qquad &   $4.56 \pm 0.09$ \\ \hline
\end{tabular}
\vspace{-.1cm}
\label{tab:mos_speech}
\end{table*}

We report the 5-scale Mean Opinion Score~(MOS), synthesis speed and model footprint in Table~\ref{tab:mos_speech}. We draw the following observations:
\begin{itemize}[itemsep=0.00pt,topsep=0pt,,leftmargin=1.5em]
\vspace{-.15cm}
\item The small WaveFlow~(res. channels 64) has 5.91M parameters, and can synthesize 22.05 kHz high-fidelity speech~(MOS: $4.32$) 42.60$\times$ faster than real-time.
In contrast, the speech quality of small WaveGlow~(res. channels 64) is significantly worse~(MOS: $2.17$). Indeed, WaveGlow~(res. channels 256) requires 87.88M parameters for generating high-fidelity speech.
\vspace{-.12cm}
\item The large WaveFlow~(res. channels 256) outperforms the same size WaveGlow in terms of speech fidelity~(MOS: $4.43$ vs. $4.34$).
It also matches the state-of-the-art WaveNet while generating speech $8.42\times$ faster than real-time, because it only requires 128 sequential steps~(number of flows $\times$ height $h$) to synthesize very long waveforms with hundreds of thousands time-steps.
\vspace{-.15cm}
\item ClariNet has the smallest footprint and provides reasonably good speech fidelity~(MOS: 4.22) because of its ``mode seeking'' behavior. 
In contrast, likelihood-based models are forced to model all possible variations existing in the data, which can lead to higher fidelity samples as long as they have enough model capacity.
\vspace{-.1cm}
\end{itemize}

\begin{figure}[t!] \centering
\vspace{-.55cm}
\begin{tabular}{cc}
\hspace{-.4cm}
\includegraphics[height=3.4cm, clip]{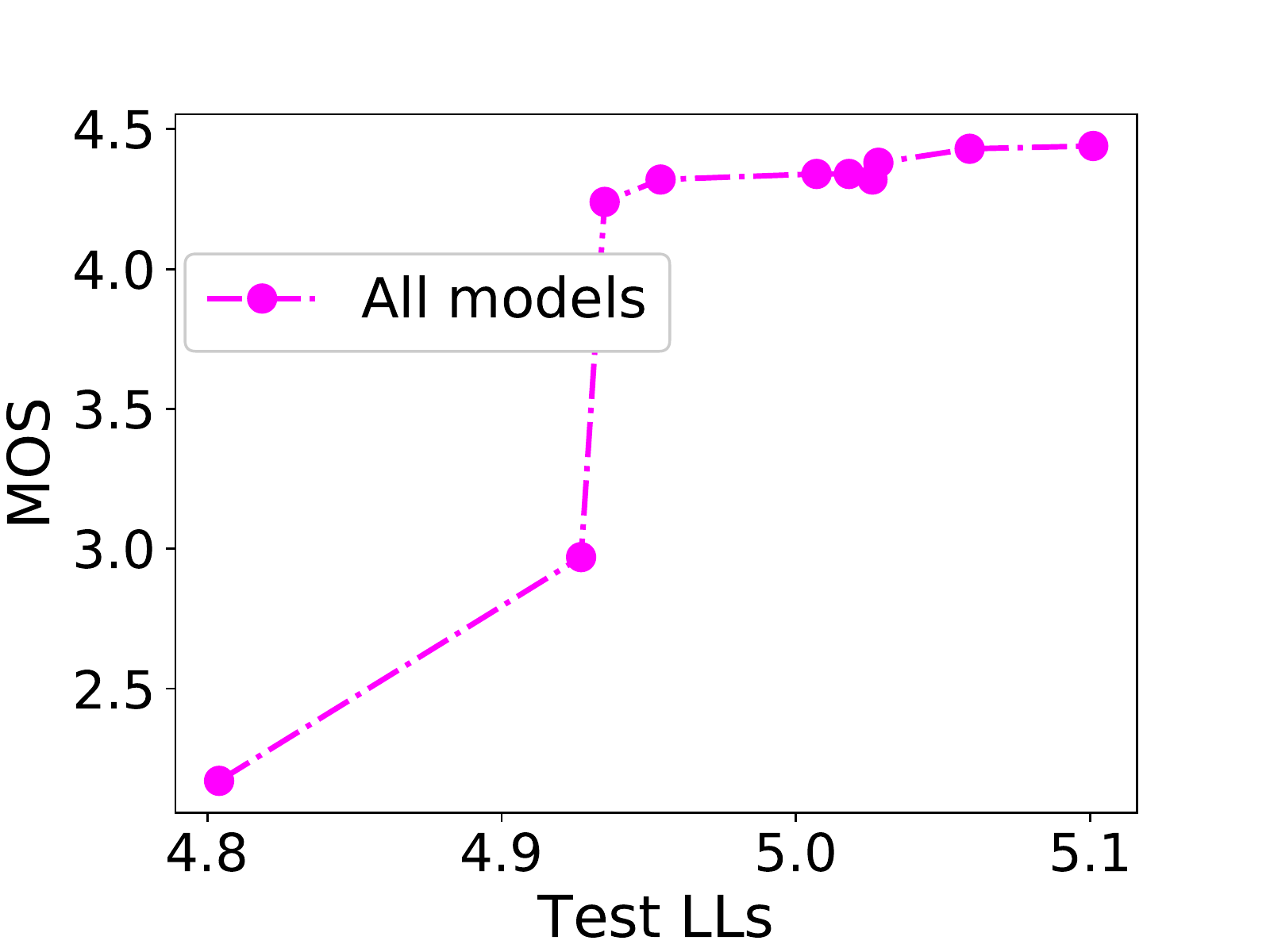} &
\hspace{-.7cm}
\includegraphics[height=3.4cm, clip]{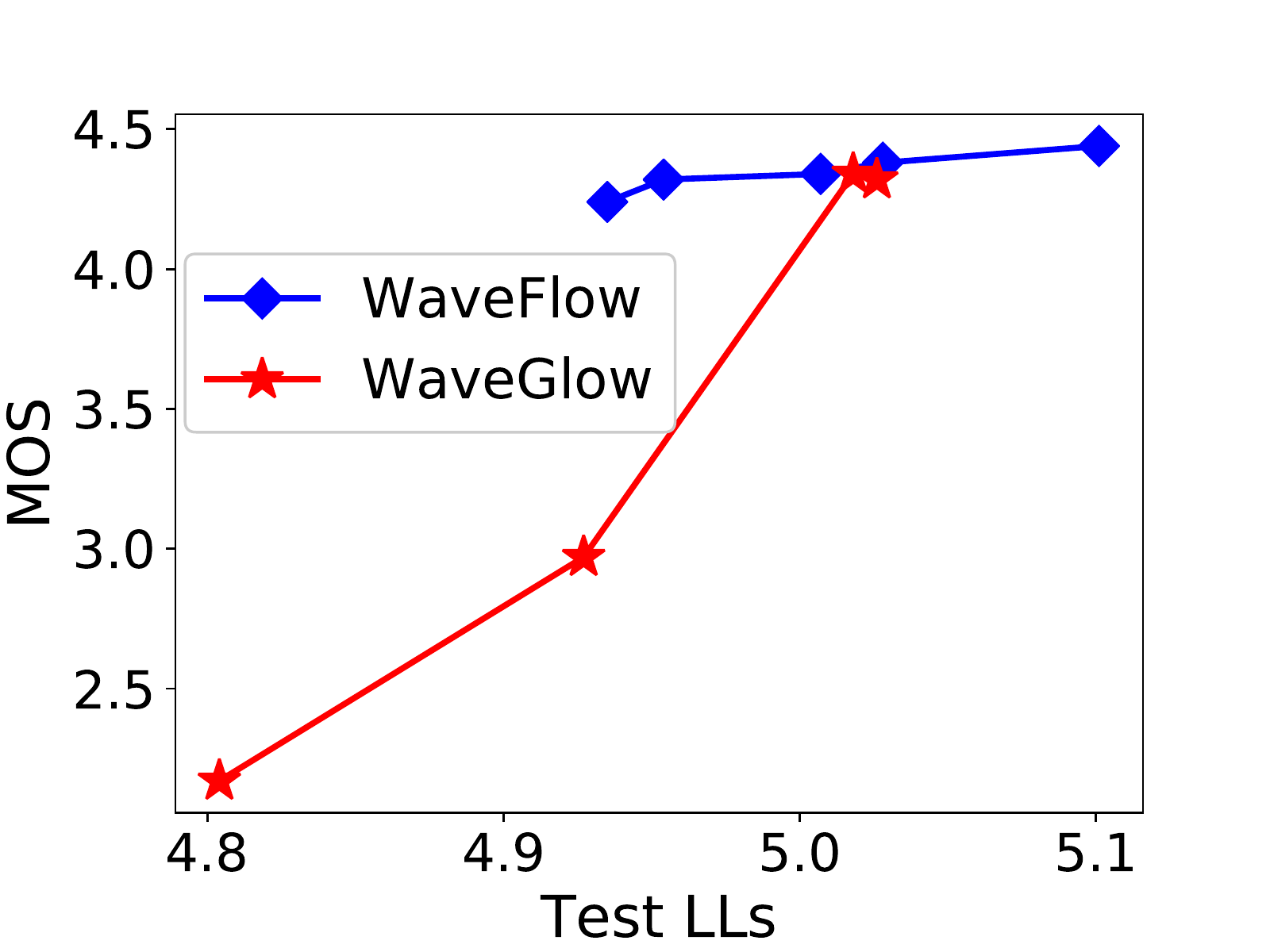} \\
\end{tabular}
\vspace{-.45cm}
\caption{The test log-likelihoods~(LLs) vs. MOS scores for all likelihood-based models in Table~\ref{tab:mos_speech}.}
\vspace{-.2cm}
\label{fig:testLLs_vs_MOS} %
\end{figure}
We also note a positive correlation between the test likelihoods and MOS scores for likelihood-based models~(see Figure~\ref{fig:testLLs_vs_MOS}).
The larger LLs roughly correspond to higher MOS scores even when we compare all models.
This correlation becomes even more evident when we consider each model separately.
It suggests that one may use the likelihood score as an objective measure for model selection.

\vspace{-.2em}
\subsection{Text-to-Speech}
\vspace{-.1em}
We also test WaveFlow for text-to-speech on a proprietary dataset for convenient reasons. It contains 20 hours of audio from a female speaker with a sampling rate of 24 kHz. We use Deep Voice 3~(DV3)~\citep{ping2017deep} to predict mel spectrograms from text. We train a 20-layer WaveNet (res. channel = 256, \#~param = {9.08~M}),~\footnote{The WaveNet hyperparameters were tuned for internal data.} 
WaveGlow (\#~param = {87.88~M}), and WaveFlow~($h$ = 16, \#~param = {5.91~M}) which are conditioned on teacher-forced mel spectrograms from DV3.
For WaveGlow, we apply the denoising function with strength 0.1 in the repository to alleviate the constant frequency noise in synthesized audio.
For WaveFlow, we sample $Z$ from isotropic Gaussian with standard deviation 0.95 to counteract the mismatch of mel conditioners between teacher-forced training and autoregressive inference from DV3.
We report the MOS results in Table~\ref{tab:tts-mos}.
%
%
As a result, WaveFlow is a very compelling neural vocoder, which features \emph{i}) simple likelihood-based training, \emph{ii}) high-fidelity \& ultra-fast synthesis, and \emph{iii}) small memory footprint.

\begin{table}[t]
\centering
\vspace{-1.2em}
\caption{MOS ratings with 95\% confidence intervals in text-to-speech experiments.}
\label{tab:tts-mos}
\vspace{0.2em}
\begin{tabular}{l|c}
\hline
\textbf{ Method }  & \textbf{MOS} \\ \hline
Deep Voice 3 + WaveNet & $ 4.21 \pm 0.08$  \\ 
Deep Voice 3 + WaveGlow  & $ 3.98 \pm 0.11$  \\ 
Deep Voice 3 + WaveFlow  & $ 4.17 \pm 0.09$  \\    \hline
\end{tabular}
\vspace{-1.0em}
\end{table}

\vspace{-.2em}
\section{Discussion}
\label{sec:conclusion}
\vspace{-.1em}
Parallel WaveNet and ClariNet minimize the reverse KL divergence~(KLD) between the student and teacher models in probability density distillation, which has the ``mode seeking'' behavior and leads to whisper voices in practice. 
As a result, several auxiliary losses are introduced to alleviate the problem, including STFT loss, perceptual loss, contrastive loss and adversarial loss~\citep{oord2017parallel, ping2018clarinet, wang2019neural, yamamoto2019probability}. In practice, it complicates the system tuning and increases the cost of development.
Since it does not need to model the numerous modes in real data distribution, a small-footprint model can generate good quality speech, when the auxiliary losses are carefully tuned.
It is worth mentioning that GAN-based models also exhibit similar  ``mode seeking'' behavior for speech synthesis~\citep{kumar2019melgan, binkowski2019high, yamamoto2019parallel}. 
In contrast, likelihood-based models~(WaveFlow, WaveGlowl WaveNet) minimize the forward KLD between the model and data distribution.
Because the model is forced to learn all possible modes within the real data, the synthesized audio can be very realistic with enough model capacity. 
However, when the model does not have enough capacity, its performance degrades quickly due to the ``mean seeking'' behavior of forward KLD~(e.g., WaveGlow with 128 res. channels).

Although audio signals are mostly dominated by low-frequency components~(e.g., in terms of amplitude), human ears are very sensitive to high-frequency content.
As a result, it is crucial to accurately model the local variations of waveform for high-fidelity synthesis, which is indeed the strength of autoregressive models.
However, autoregressive models are less efficient at modeling long-range correlations, which can be seen from the difficulties to generate globally consistent images~\citep{van2016conditional, menick2018generating}. 
Worse still, they are also noticeably slow at synthesis.
Non-autoregressive convolutional architectures can do speedy synthesis and easily capture the long-range structure in the data~\citep{radford2015unsupervised, brock2018large}, but it could produce spurious high-frequency components which will hurt the audio fidelity~\citep[e.g.,][]{donahue2018adversarial}.
In this work, WaveFlow compactly models the local variations using short-range autoregressive functions, and handles the long-range correlations with a non-autoregressive convolutional architecture, which obtains the best of both worlds.

\nocite{wang2019neural}

\bibliography{reference}
\bibliographystyle{icml2020}

\onecolumn

\appendix

\section*{{\Large Appendix}}

\section{Squeezing time-domain samples on channel dimension may raise artifacts}
\label{appendix:waveglow_noise}
\begin{figure}[h] \centering
\begin{tabular}{c}
\hspace{-.3cm}
\includegraphics[height=2.8cm, clip]{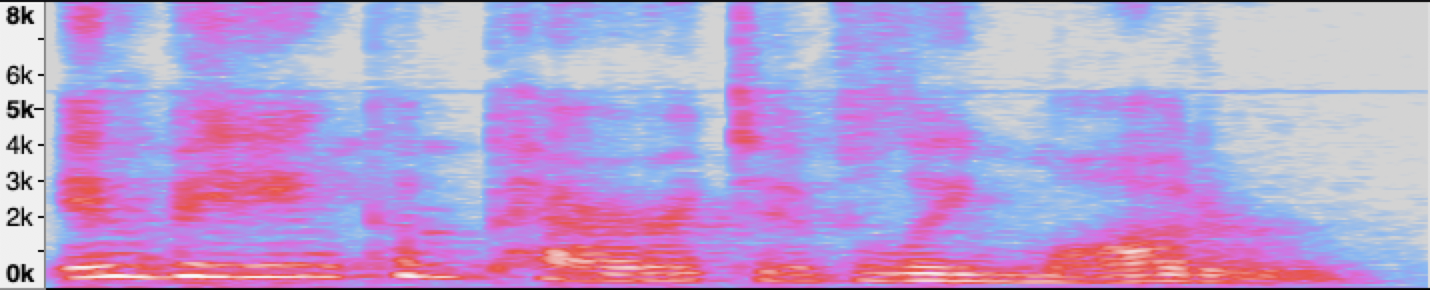} 
\\
{\small (a) squeezing size = 4. The frequency of noise is $\frac{22.05}{4} \approx 5.5$ kHz (the horizontal line).}
\vspace{0.2cm}
\\
\hspace{-.3cm}
\includegraphics[height=2.8cm, clip]{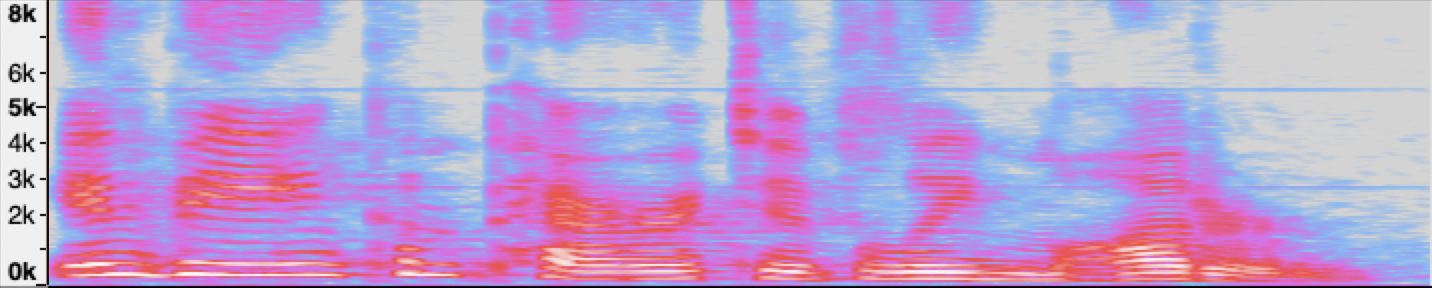} 
\\
{\small (b) squeezing size = 8. The frequencies of the noises are $\frac{22.05}{4}=5.5$ kHz and $\frac{22.05}{8}\approx 2.75$ kHz.}
\vspace{-.5em}
\\
\end{tabular}
\vspace{0.2em}
\caption{(best viewed in color) The spectrograms of synthesized audios from WaveGlow by setting (a) squeezing size = 4 without early output, and (b) squeezing size = 8 with early output (default in the WaveGlow repository).}
\label{fig:waveglow_noise} %
\end{figure}

It is inefficient for modeling raw waveform by squeezing the time-domain samples on channel dimension and applying feed-forward transformation, because one can lose the temporal order information within the squeezed sequence.
As a result of doing this, the synthesized audios from WaveGlow may contain constant frequency noise~(see Figure~\ref{fig:waveglow_noise}).
Note that, the frequencies of the noises (horizontal lines on spectrogram) are directly related to the squeezing size.  

\end{document}